\documentclass[a4paper,11pt]{article}
\usepackage{pos}

\title{Baryonic Axion in neutron-antineutron oscillation}

\author*[a]{Théo Brugeat}

\affiliation[a]{Laboratoire de Physique Subatomique et Cosmologie,\\
  53 Avenue des Martyrs, 38000 Grenoble, France}

\emailAdd{theo.brugeat@lpsc.in2p3.fr}

\abstract{The accidental baryonic symmetry is expected to be broken and required from the observed matter-antimatter asymmetry. The neutron-antineutron oscillating system is the hallmark of $\Delta \mathcal{B}=2$ models which have the benefits of not inducing proton decay. We study this system in a framework allowing the most general couplings to understand how a dark matter candidate such as the axion may couple to the oscillation. In particular a Rabi resonance phenomenon occurs, and this effect is unconstrained for Axion Like Particles (ALPs) models. Regarding the QCD axion, its Goldstone nature leads to a robust exclusion of the majority of scenarios allowed.}

\FullConference{2nd Training School and General Meeting of the COST Action COSMIC WISPers  (CA21106) (COSMICWISPers2024)\\
 10-14 June 2024 and 3-6 September 2024\\
Ljubljana (Slovenia) and Istanbul (Turkey)\\}

\begin{document}
\maketitle

\section{Introduction}
The accidental $U(1)_{\mathcal{B}}$ symmetry of the Standard Model is supposed to be broken by higher dimensions operators, the model allows for two $(\Delta\mathcal{B}=2)$ quark's operators of dimension 9. These couplings are thus expected to be suppressed by a factor $(\Lambda_{QCD}/\Lambda_{\Delta\mathcal{B}=2})^{5}%
\sim10^{-18}~(1~$TeV$/\Lambda_{\Delta\mathcal{B}=2})^{5}$ using $\Lambda
_{QCD}\approx300~$MeV. The standard effective \cite{Mohapatra:2009wp} approach to describe the system of neutron oscillation consider the neutral Dirac fermion with a weak coupling to the proton, a magnetic dipole moment and a small Majorana mass term:
\begin{equation}\label{Standard}
		\mathcal{L}= \bar{n}(i \!\not\! \partial-m_{D})n - \varepsilon_0 \left(\bar{n}n^\mathcal{C} + \bar{n}^\mathcal{C}n\right) - \frac{\mu}{2} \bar{n}\sigma_{\mu\nu} n F^{\mu\nu}.
\end{equation}
In the non-relativistic approximation, the dynamic holds in a system of coupled Schrödinger equations, its resolution leads to the following transition probability of the neutron into an antineutron:
\begin{equation}
    P_{n\rightarrow n^\mathcal{C}}(t) = e^{-\Gamma t} \frac{\varepsilon_0^2}{\left(\Delta E/2\right)^2 +\varepsilon_0^2}\sin^2\left( \sqrt{\left(\Delta E/2\right)^2 +\varepsilon_0^2} t\right)
\end{equation}
The energy splitting $\Delta E$ is proportional to the magnetic field, the current bound is measured in the quasi-free regime where $B\rightarrow 0$ leading to $ P_{n\rightarrow n^\mathcal{C}}(t) \simeq \varepsilon_0^2 t^2$, the constraint on the Majorana parameter is strong, accordingly to the scale dependence of the quarks operators involved: $\varepsilon_0<0.8\times10^{-23}~$eV. The motivation for the present work and the associated article lies in the fact that promoting the parameter $\varepsilon(t) \rightarrow \varepsilon_0 \sin{(\omega t)}$ leads to a new transition probability which features a Rabi resonance which may allow to significantly increase the signal regardless of $\varepsilon_0$:
\begin{equation}\label{Rabi}
    P_{n\rightarrow n^\mathcal{C}}(t) = e^{-\Gamma t} \frac{\varepsilon_0^2}{\left((\omega - \Delta E)/2\right)^2 +\varepsilon_0^2}\sin^2\left( \sqrt{\left((\omega - \Delta E)/2\right)^2 +\varepsilon_0^2}t \right)
\end{equation}
Such dynamical coupling may arise from a dark sector, in our case axions and ALPs are favored candidates. Behind the scene, a treatment of the general setup is required including all the couplings allowed in the $\Delta \mathcal{B} = 2 $ model. To realize such scenario with $\epsilon(t)$ the couplings, $\phi\bar{n}^\mathcal{C}n$, $a\bar{n}^\mathcal{C}\gamma^5n$ or $\partial_\mu a\bar{n}^\mathcal{C}\gamma^\mu\gamma^5n$ can be used, and we considered the last two. As an outcome, the QCD axion is almost excluded, but ALPs remain unconstrained. 
\section{Baryonic Axion Models}
The UV coupling of a QCD axion to the baryonic sector is achieved by identifying the Baryonic $U(1)_\mathcal{B}$ to  the Peccei-Quinn symmetry \cite{Arias-Aragon:2022byr}. The Goldstone nature of the axion then imposes that its linearized couplings $a\bar{n}^\mathcal{C}\gamma^5n$ appears as the expansion of the exponentiated term induced by the Peccei-Quinn scalar field, $\exp{(ia/v)}\bar{n}^\mathcal{C}n$, with $v$ the Peccei-Quinn scale. The reparametrization invariance allows to render the shift invariance explicit and leaves parametric Majorana masses in addition to the derivative coupling with the baryonic current and to the coupling with its anomaly:
\begin{equation}
    \mathcal{L}=\frac{1}{2}\partial_\mu a\partial^\mu a + \bar{n}(i \!\not\! \partial-m_{D})n + \partial_\mu a \bar{n}\gamma^\mu n - \left( m_L \bar{n}^\mathcal{C}n + m_R \bar{n}n^\mathcal{C} + h.c. \right) + \delta \mathcal{L}_{Jac}
\end{equation}
The coupling to gluons required to solve strong CP is separately induced, e.g. via heavy colored fermions
à la KSVZ. The Majorana masses indicate that the neutrons are not the eigenstates, hence the baryonic current is no longer conserved, and consequently a coupling to the Baryonic violating currents occurs. It can be seen if one transforms the fermionic fields to realize \eqref{Standard}, as the transformation will mix the baryonic current with $\partial_\mu a\bar{n}^\mathcal{C}\gamma^\mu\gamma^5n$.
\section{Reduction of the general Lagrangian into the standard system}
The mass sector can be implemented into a matrix to recast the Lagrangian with the two Weyl fermions of the system:
\begin{equation}
     M =\left[\begin{array}{cc}
		m_L & m_D  \\
		m_D & m_R \\
	\end{array}\right]
\end{equation}
The kinetic sector being invariant under unitary mixing of the two fermions in this formulation, a $U(2)$ matrix can always be used to transform the system into \eqref{Standard}, to analyze the oscillation \cite{Berezhiani:2018xsx,Fujikawa:2016sft}. The mass sector is transforming as $M\rightarrow U^T\cdot M\cdot U$ under the unitary matrices which consist of Chiral and baryonic rephasings, and of Bogolyubov rotations mixing $n$ and $n^\mathcal{C}$. The way $M$ transforms is known in the case of neutrinos but the explicit form of U to transform a general mass matrix into \eqref{Standard} is quite complicated, see \cite{Brugeat:2024rxe}. Notice that the oscillation can be turned off in the limiting case $m_L=-m_R^*$ but this requires $U\neq1$ to obtain $\varepsilon_0 = 0$. Under a transformation $U \in U(2)$ every non-kinetic sector of the Lagrangian is affected differently, and we can mention some effects:
\begin{itemize}
    \item The magnetic and electric dipole moments are mixed as $\mu \rightarrow \mu \det{U}$ and thus only the chiral $U(1)$ contributes, this generalizes the standard requirement that $m_D$ and $\mu$ have their phase aligned, to suppress the neutron electric dipole moment generated by a misalignment.
    \item The Weak sector is extended with $\Delta \mathcal{B}=2$ couplings which leads to mixing without oscillation. The axial and vector couplings $g_A,g_V$ are mixed as well because of $U(1)$.
    \item The baryonic current $J_3^\mu=\bar{n}\gamma^\mu n$ gets mixed with the $\Delta \mathcal{B}=2$, CP-conserving and CP-violating currents $J_1^\mu=(\bar{n}\gamma^\mu \gamma^5 n^\mathcal{C}+\bar{n}^\mathcal{C}\gamma^\mu \gamma^5 n )$ and $J_2^\mu=i(\bar{n}\gamma^\mu \gamma^5 n^\mathcal{C}-\bar{n}^\mathcal{C}\gamma^\mu \gamma^5 n )$ respectively. A natural description of these mixings lies in the transformation of $\Vec{J}$ in the $SO(3)$ representation associated to $U(2)$.
    \item All three fore-mentioned currents becomes anomalous in general once we reach the standard basis, this can be understood from the $\Delta \mathcal{B}=2$ weak couplings.
\end{itemize}
\section{Axionic $\Delta\mathcal{B}=2$ effects}
After the reparametrization, the Baryonic current coupled to the axion is mixed by the unitary transformation, the leading order can be deduced in our case as:
\begin{equation}
        \partial_\mu J_3^\mu \rightarrow \partial_\mu \left(r_1 J_1^\mu +r_2 J_2^\mu+J_3^\mu \right) + O(\epsilon / m_D),
\end{equation}
with $r_1 = O(m_{L/R}/m_D)$ and $r_2 = O(m_{L/R}\sin{\phi}/m_D)$ and $\phi$ a combination of the phases involved initially in the mass matrix. The mixing of the currents leads to an unusual formulation of the equivalence theorem $a \bar{n}^\mathcal{C}\gamma^5n \leftrightarrow \partial_\mu a \bar{n}^\mathcal{C} \gamma^\mu\gamma^5n $, in terms of Ward identities. In the standard basis, the baryonic current being not Baryon-violating cannot induce an oscillation, but $\partial_\mu a J_{1,2}^\mu$ coupling are good candidates and their non-relativistic contribution reduces to:
\begin{equation}
    \epsilon(t) \simeq \epsilon + \frac{r_{1,2}}{v} \sigma \cdot  \nabla a 
\end{equation}
The gradient implies a reduction by the Dark matter velocity  if the axion is chosen as a DM candidate. The oscillation induced by $\varepsilon$ is much bigger and already constrained, and if one can set $\varepsilon =0$, the mixing produced by $\Delta \mathcal{B} = 2$ weak couplings constrains $r_{1,2}$. Consequently, the QCD-axion cannot produce a significant oscillation. This is bypassed by ALP's which doesn't fulfill the equivalence theorem allowing to impose $\varepsilon=0$ and keep the shift-symmetric coupling only.

\section{Conclusion}
To conclude, we have investigated the scenario of baryonic dark matter with a focus on QCD axions and ALPs, and we suggest that a resonance phenomenon may happen from a scalar coupling to the $n-\bar{n}$ oscillating system.
\begin{itemize}
    \item For QCD axions, which are Goldstone modes satisfying the equivalence theorem, we concluded that their nature makes them irrelevant phenomenologically regarding the $n-\bar{n}$ oscillation.
    \item We identified the leading contribution of axionic couplings to the oscillating $n-\bar{n}$ system in a general setup. We reach the Rabi resonance we aimed at \eqref{Rabi} but the latter has an extra derivative than expected, proportional to the axion-wind: $\varepsilon(t)\propto \frac{\nabla a}{2v}$.
    \item Axion like particles are not constrained as much as QCD axions as they can be coupled to $\Delta \mathcal{B}=2$ sector through only a derivative coupling whose coupling constant is then a free parameter of the model. This kind of baryonic dark matter is a new sector beyond standard model that is worth some further investigations.
    \item This work is transposable to the case of neutrinos for which the leptonic current may be coupled to a majoron identified to the axion, although the flavor symmetry is enlarged, the formalism hold and we expect that the diagonalization of the neutrino mass would mix the currents coupled to the Goldstone mode. One significant difference in this scenario is the reversed mass scaling between Dirac and Majorana terms which may reveal a different phenomenology once expanded. 
\end{itemize}

\end{document}